# Discovery of an Arc System in the Brightest ROSAT Cluster of Galaxies

S. Schindler[1,2], L. Guzzo[3], H. Ebeling[4], H. Böhringer[1], G. Chincarini[3,5], C.A. Collins[6], S. De Grandi[5], D.M. Neumann[1], U.G. Briel[1], P. Shaver[7], G. Vettolani[8]

[1] Max–Planck–Institut für extraterrestrische Physik, Giessenbachstraße, D–85478 Garching, Germany
[2] Max–Planck–Institut für Astrophysik, Karl–Schwarzschild–Straße 1, D–85478 Garching, Germany
[3] Osservatorio Astronomico di Brera, Via Bianchi 46, I–22055 Merate (CO), Italy
[4] Institute of Astronomy, Madingley Road, Cambridge CB3 0HA, UK
[5] Dipartimento di Fisica, Università di Milano, Via Celoria 16, I–20133 Milano, Italy
[6] School of Chemical and Physical Sciences, Liverpool John–Moores University, Byrom Street, Liverpool L3 3AF, UK
[7] ESO, Karl–Schwarzschild–Straße 2, D–85748 Garching, Germany
[8] Istituto di Radioastronomia del CNR, Via Gobetti 101, I–40129 Bologna, Italy

[the date of receipt and acceptance should be inserted later]

**Abstract.** We report the discovery of two bright arcs in what turns out to be the brightest X–ray cluster in the ROSAT band ever observed, RXJ1347.5-1145. Its luminosity is $(6.2 \pm 0.6) \cdot 10^{45}$ erg s$^{-1}$ (in the range 0.1–2.4 keV). The arcs are most probably gravitationally lensed images of background galaxies. They were found serendipitously during our ongoing large–scale redshift survey of X–ray clusters detected by the ROSAT All Sky Survey. The arcs are almost opposite to each other with respect to the cluster centre, with a distance from it of about $35''$ (= $240 h_{50}^{-1}$ kpc), a radius that enables the probing of a rather large cluster volume. In this Letter we limit ourselves to the discussion of the general optical and X–ray features of this cluster and to the potential implications of the gravitational arcs. A more detailed discussion of the different mass estimates and of the cosmological implications for this exceptional object are left for future work based on more accurate optical and X–ray data, which are currently being collected.

## 1. Introduction

The deepest, large–scale gravitational potentials in the universe are found in the centres of galaxy clusters. These huge gravitational fields deflect electromagnetic waves and can therefore act as giant gravitational lenses distorting and enhancing the images of background galaxies (Lynds & Petrosian 1986; Soucail et al. 1987). Up to now a few tens of clusters of galaxies with luminous arcs have been found (Refsdal & Surdej 1994; Fort & Mellier 1994). These arcs are interesting in two respects. Firstly, the magnification through gravitational lensing enables the investigation of distant galaxies, which would otherwise be too faint for spectroscopic studies. Secondly, arcs can be used to model the cluster potential and allow the total mass of the cluster within the arcs to be estimated. A good example is the cluster A370 for which Kneib et al. (1993) reconstructed a very detailed model potential and lens configuration from the observations of a giant arc and three arclet pairs. A mass estimate for the cluster can also be derived through X–ray observations of the intracluster gas assuming hydrostatic equilibrium. Interestingly, for the few well observed examples analysed so far the masses derived in the two ways differ by up to a factor of three with the mass from gravitational lensing being usually the larger one (see e.g. Miralda–Escudé & Babul 1994). It is therefore of particular interest to find and study new systems of gravitational arcs in clusters that are clearly detected in the X–rays.

A byproduct of the redshift survey of ROSAT clusters that we are performing over the whole southern sky as an ESO key–programme (Böhringer 1994; Guzzo et al. 1995), are short direct images of the candidates that are routinely carried out in preparation for the spectroscopy. Here we would like to report on the detection during one of these observations of a system of *bona fide* bright arcs, related to a rather interesting cluster identified with the X–ray source RXJ1347.5-1145. Despite its anonymity [the two previous similar discoveries related to ROSAT clusters corresponded to known optically selected clusters, S295 (Edge et al. 1994) and A2104 (Pierre et al 1994)],

---

at a redshift $z = 0.451$ this object turns out to be the most X–ray luminous cluster known. While we have already started a systematic campaign of deep optical and X–ray pointed observations, we found it important to publish in this Letter a short account on the general properties of this exceptional object, with special emphasis on the implications of the gravitational arcs. Future work, based on the more accurate new data, will be more focused on the mass estimates and on the cosmological relevance of the very existence of such a massive cluster.

Throughout the paper we use a Hubble constant of $H_0 = 50$ km s$^{-1}$Mpc$^{-1}$.

## 2. Optical properties

In Fig.1 we show the central part (1.4 arcmin) of a 10-minute R exposure of the field of the ROSAT source RXJ1347.5-1145, obtained at the ESO 2.2 m telescope using EFOSC II (Buzzoni et al. 1984). A very rich cluster of galaxies with two dominant galaxies is centred on the position of the X–ray source. Both arcs are located at about 35 arcseconds from the cluster centre and are positioned on opposite sides of the central galaxy. They are aligned perpendicularly to the radial direction and the position angle of the line connecting the two arcs is 36° (North to East). Nearly in the middle between the two arcs is the brightest galaxy of the cluster at the position $\alpha = 13^h47^m30.5^s\ \delta = -11°45'09''$ (2000). The arcs are $5''$ and $6''$ long, their positions are given in Table 1. The northern arc has a slightly brighter spot on the eastern end. From our data it is not clear whether this spot is really part of the arc or a foreground source. If it is a foreground source the length of the northern arc is between $4''$ and $6''$. The ratio of the length to the width of the arcs is large ($\geq 4$) although the arcs could not be resolved with the seeing of about $1''$. A possible smaller third arc is in the southeast with a distance from the centre of 45 arcseconds (see Table 1). It also has an orthoradial orientation relative to the central galaxy.

Although we do not yet have spectroscopic confirmation of the arcs, there is convincing evidence that they cannot be edge–on spiral galaxies. None of the arcs exhibit a bulge or a bright nucleus, while their luminosity is not compatible with being irregular cluster members. Indeed, we can estimate from simple photon counting the difference between the brightest cluster galaxy and the arcs to be 2.5 - 3 magnitudes. The first-ranked galaxies in a cluster can be taken roughly as standard candles (Sandage & Hardy 1973). The luminosity function of cluster galaxies (e.g. Binggeli et al. 1988) shows that the brightest irregular galaxies are at least 5 magnitudes fainter than the brightest cluster galaxy.

In Fig.2 we show the spectra of the two brightest cluster members, observed at low resolution ($\sim 1000$ km s$^{-1}$) with EFOSC II in spectroscopic mode. A number of features are evident in both spectra. The central galaxy shows both absorption and emission lines, indicating the presence of both an old stellar population and active star formation (and/or nuclear activity). From the direct image one can also notice the presence of a number of smaller very close objects around this galaxy (note also the outer region devoid of any galaxy). The other, eastern dominant galaxy shows a classical early–type absorption spectrum. The redshifts of the two galaxies, obtained using all the available absorption/emission features, are in very good agreement and once combined provide a mean value for the cluster of $z = 0.451 \pm 0.003$.

**Table 1.** Positions and lengths of the arcs in RXJ1347.5-1145

|  | $\alpha$(2000) | $\delta$(2000) | length |
|---|---|---|---|
| Southern arc | $13^h47^m29.2^s$ | $-11°45'39''$ | $5''$ |
| Northern arc | $13^h47^m32.0^s$ | $-11°44'42''$ | $6''$ |
| Possible arc | $13^h47^m31.8^s$ | $-11°45'50''$ | $3''$ |

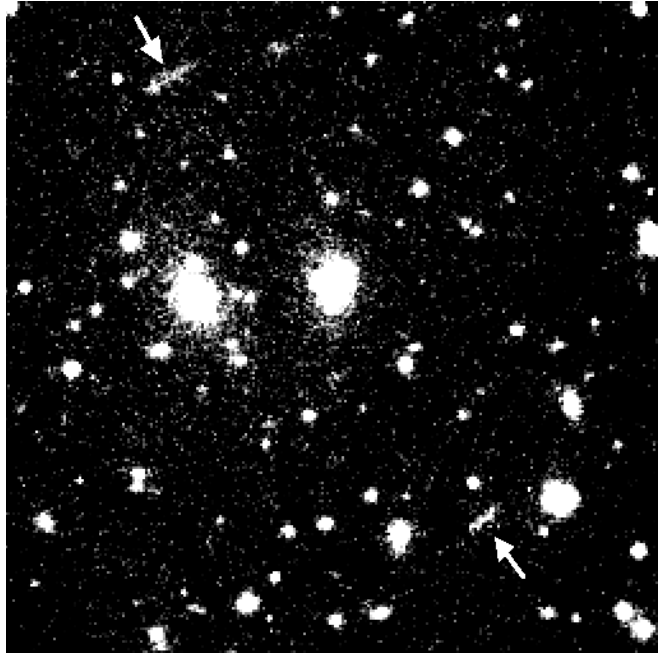

**Fig. 1.** R image of the centre of the cluster RXJ1347.5-1145 over a square field with 1.4 arcmin side slightly smoothed with a Gaussian with $\sigma$=0.5pixel. The two arcs are northeast and southwest of the central galaxy (North is up, East is left).

## 3. X–ray properties

The source RXJ1347.5-1145 was detected in the ROSAT All–Sky Survey (RASS) with 118 source photons in the broad ROSAT band (0.1-2.4 keV) and 104 source photons in the hard band (0.5-2.0 keV) with an exposure time of 370 sec. For a range of temperatures between 4 and 12 keV we derive an X–ray luminosity of $(6.2 \pm 0.6) \cdot 10^{45}$erg s$^{-1}$ (0.1-2.4 keV). This luminosity makes it the brightest cluster in the broad ROSAT band yet discovered, even exceeding the luminosity of A2163 (Elbaz et al. 1995). The cluster was detected as an extended X–ray source by the source detection algorithm of the standard analysis applied to the RASS. The X–ray emission can be traced out to a radius of $5.5'$ ($1.8 h_{50}^{-1}$Mpc). Fig. 3 shows a slightly smoothed image of the cluster. The cluster is elongated with a position angle of about $-45°$, that is, almost perpendicular to the line connecting the two arcs. In the outermost parts the position angle of the X–ray emission rotates to $< -60°$. While in all other arc clusters with good X–ray images the arcs lie on the major axis, it is the unique feature of this system to have the arcs on the minor axis. The cluster is a hard X–ray source with a hardness ratio which is consistent with the spectrum of X–ray emission from hot intracluster gas ($n_H = 3.58 \cdot 10^{20}$cm$^{-2}$ (Dickey & Lockman 1990)).

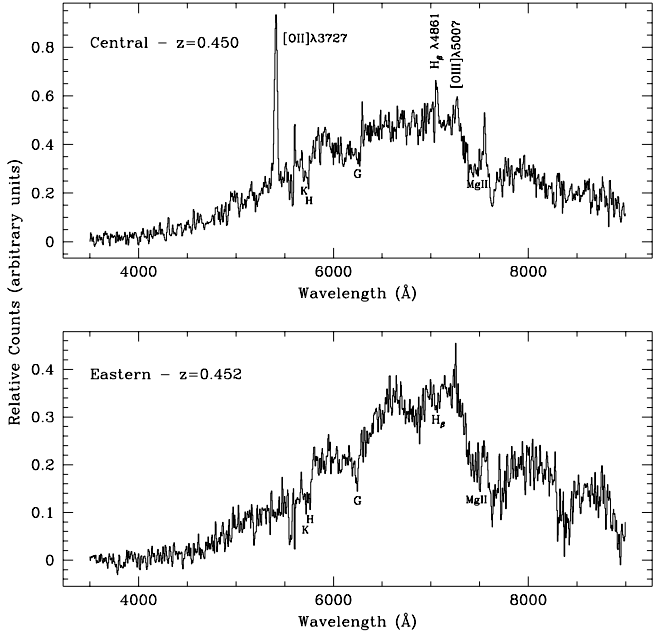

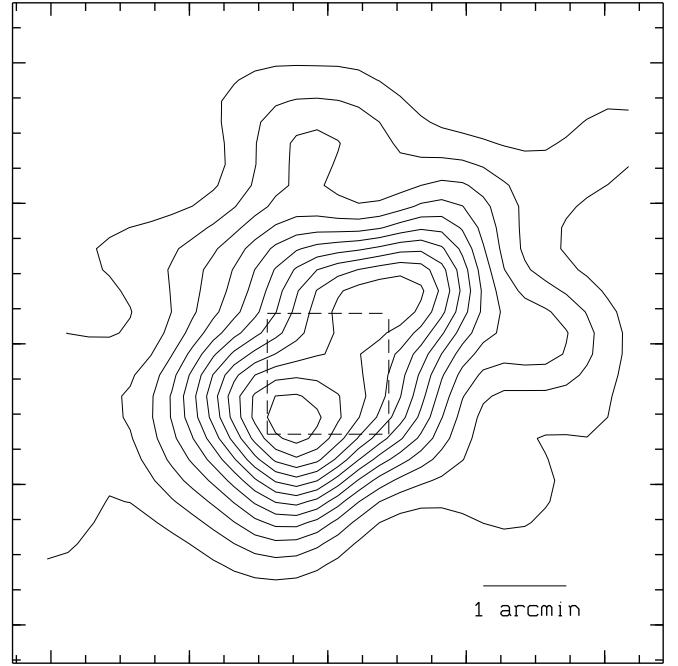

**Fig. 2.** Spectra of the two dominant galaxies in RXJ1347.5-1145. The major absorption and emission features are marked.

**Fig. 3.** X–ray image of the cluster RXJ1347.5-1145 from the ROSAT All Sky Survey in the hard band smoothed with a Gaussian filter of $\sigma = 30$ arcsec (North is up, East is left). The contours are linearly spaced with $\Delta$countrate = $2.4 \cdot 10^{-3}$ cts s$^{-1}$arcmin$^{-2}$ the highest contour line corresponding to $2.6 \cdot 10^{-2}$ cts s$^{-1}$arcmin$^{-2}$. The X-ray maximum is at $\alpha = 13^h 47^m 32^s$ $\delta = -11°45'42''$ (2000) (note the uncertainty of RASS source positions of $20''$). The position of Fig. 1 is indicated by the dashed box.

## 4. Mass

We can derive a crude mass estimate for the cluster by assuming a fiducial redshift for the lensed galaxies and that both arcs lay close to the Einstein ring. For a redshift $z_{gal} = 0.7 - 1.2$ (a reasonable guess based on previous observations of arcs in clusters at similar redshift), we derive a cluster mass within $240 h_{50}^{-1}$ kpc of $M_{cl} = 7.8 - 4.4 \cdot 10^{14} \mathcal{M}_\odot$, an exceptionally high value for the core region of a cluster. Note, however, that so far for most clusters only smaller radii have been probed using the gravitational lens method. The mass value we have obtained is consistent with the high X-ray luminosity of the cluster. If we assumed an elliptical mass distribution (as suggested by the X-ray image), with the arcs laying along the minor axis, we would end up at an even larger mass.

## 5. Conclusions

RXJ1347.5-1145 is the third cluster of galaxies where gravitational arcs have been serendipitously discovered during ROSAT follow–up observations, another indication of how effective X–ray selected clusters are as gravitational lens candidates. (Gioia & Luppino (1994) also found arcs in almost 30% of the Einstein Medium Sensitivity Survey clusters.) The reason for this is that X-ray emission ensures the selection of deep potential wells associated with massive and rich galaxy clusters. This is in contrast to optically selected cluster samples which can suffer from line-of-sight richness contamination. We expect, therefore, that the intrinsically luminous, high–redshift tail of the flux–limited sample selected for our ESO key–programme cluster sample will provide a homogeneous set of massive lensing clusters.

We are in the process of obtaining new X–ray observations of RXJ1347.5-1145 with the ROSAT HRI and with the ASCA satellite. These will provide a detailed picture of the intracluster gas structure and temperature and thereby of the cluster potential. On the basis of these new data, and hopefully of new optical information on the arc system, we shall be able to discuss in greater detail the cosmological implications of this extraordinary object.

*Acknowledgements.* We thank the ROSAT team for preparing the ROSAT data and for providing the analysis software. We thank J. Wambsganß, A. Buzzoni, H. Jerjen, and B. Binggeli for useful discussions. S.S. and H.B. gratefully acknowledge financial support by the Deutsche Forschungsgemeinschaft and the Verbundforschung. H.E. acknowledges support by a European Union HCM Fellowship.

## References

Binggeli B., Sandage A., Tammann G.A., 1988, ARA&A 26, 509

Böhringer H., 1994, in: Studying the Universe with Clusters of Galaxies, Böhringer H., Schindler S. (eds.), MPE Report 256, p. 93

Buzzoni B., Delabre B., Dekker H., D'odorico S., Enard D., Focardi P., Gustafsson B., Reiss R., 1984, ESO Messenger 38, 9

Dickey, J.M. and Lockman F.L., 1990, ARA&A 28, 215

Edge A.C., Böhringer H., Guzzo L., et al., 1994, A&A 289, L34

Elbaz D., Arnaud M., Böhringer H., 1995, A&A 293, 337

Fort B., Mellier Y., 1994, A&AR 5, 239

Gioia I.M., Luppino G.A., 1994, ApJS 94, 583

Guzzo L., Böhringer H., Briel U., et al., 1995, in: $35^{th}$ Herstmonceux Conference: Wide–Field Spectroscopy and the Distant Universe, Maddox, S.J., Aragón-Salamanca, A. (eds.). World Scientific, Singapore, in press

Kneib J.-P., Mellier Y., Fort B., Mathez G., 1993, A&A 273, 367

Lynds R., Petrosian V., 1986, Bull. Am. Astron. Soc. 18, 1014

Miralda-Escudé J., Babul A., 1994, preprint

Pierre M., Soucail G., Böhringer H., Sauvageot J.L., 1994, A&A 289, L37

Refsdal S., Surdej J., 1994, Rep. Prog. Phys. 56, 117

Sandage A., Hardy E., 1973, ApJ 183, 743

Soucail G., Fort B., Mellier Y., Picat J.-P., 1987, A&A 172, L14